\documentclass[a4paper,10pt]{article}
\usepackage[utf8]{inputenc}
\usepackage[T1]{fontenc}
\usepackage{graphicx}
\usepackage[a4paper,left=2cm,right=2cm,top=3cm,bottom=3cm,head=0cm,headsep=0cm]{geometry}
\usepackage{caption} 
\usepackage{subcaption}
\usepackage{amsmath}
\usepackage{amssymb}
\usepackage{authblk}
\usepackage[pagewise]{lineno}
\usepackage{hyperref}
\usepackage{lineno}
\usepackage{multirow}
\usepackage{parskip}
\usepackage{xcolor}
\usepackage[normalem]{ulem}
\usepackage{tikzsymbols}
\usepackage{rotating}
\usepackage{tabularx}
\author[1]{M.~Pietrow\thanks{Corresponding Author: marek.pietrow@umcs.pl}}
\author[2]{A.~Miaskowski}
\title{Artificial neural network as an effective tool to calculate parameters of positron annihilation lifetime spectra}
\affil[1]{Institute of Physics, M.~Curie-Sk\l{}odowska University, Pl. M.~Curie-Skłodowskiej 1, 20-031 Lublin, Poland}
\affil[2]{Faculty of Production Engineering, University of Life Sciences, Akademicka 13, 20-950 Lublin, Poland}

\date{\today}
\begin{document}
\maketitle
\begin{abstract}
The paper presents the application of the multi-layer perceptron regressor model for predicting the parameters of positron annihilation lifetime spectra using the example of alkanes in the solid phase. A good agreement of calculation results was found when comparing with the commonly used methods. The presented method can be used as an alternative quick and accurate tool for decomposition of PALS spectra in general. The advantages and disadvantages of the new method are discussed.
\end{abstract}
\section{Introduction}
Positron Annihilation Lifetime Spectroscopy (PALS) is one of the useful experimental methods using positrons for studying structural details in a wide spectrum of materials, in particular in the solid state~\cite{Piro20}. This method is based on the annihilation of positrons where their lifetime and annihilation intensity in the sample is dependent on some properties of the material in the nano scale, including local electron density, bound electron energy, and the density and size of free volumes in the sample.
\par
Depending on the material, besides the process of direct annihilation, a positron can form a meta-stable atomic state with an electron, called a positronium (Ps), which can exist in two spin states referred to as \emph{para}- and \emph{ortho}-Ps differing in properties (especially, their lifetimes differ in vacuum by three orders of magnitude)~\cite{GreinerField}. A number of conditions must be met for the Ps to be formed in matter. One of them is that free volumes of a sufficiently large size must be present. For these materials, a Ps is extremely useful in material science since its lifetime can be related to the size of free volumes~\cite{Goworek15}. Depending on the structure of the sample, there is possibly a variety of Ps components which annihilate with characteristic lifetimes. All these populations give their own account to the positron annihilation spectrum measured experimentally. PALS spectra require decomposition in the post-measuring procedure of decomposition resulting in both the calculation of lifetimes for particular species of positrons and the relative amplitudes for these processes (so-called spectrum inversion problem)~\cite{Jean13}.
\par
Many algorithms used for data processing require assuming an exponential character of positron decay. They also require fixing the number of components used during the decomposition. For example, the method used by one of the adequate software, the LT programme~\cite{Kansy96} or PALSfit~\cite{PALSfit07}, consists in fitting the PALS experimental spectrum to a sum of a given number of exponential functions usually convoluted with the (multi) gaussian apparatus resolution curve.
\par
The PALS spectra used here were measured for normal alkanes (n-alkanes), i.e. the simplest organic molecules where carbon atoms form a straight chain of the molecule and are saturated by hydrogen atoms. The n-alkanes with a different number $n$ of carbon atoms in the molecule form a homologous series described by the general chemical formula C$_n$H$_{2n+2}$ (C$n$ is used as an abbreviation). Alkanes in the solid phase form molecular crystals where the trains of elongated molecules are separated by gaps called the inter-lamellar gaps. Ps can be formed in the free volumes made by both the gaps and the spaces generated by changes in the conformation with temperature~\cite{Goworek15}. Using the PALS technique, the size of these free volumes can be determined from the lifetime and the relation between both being given by the Tao-Eldrup formula or its modification~\cite{Wada13}. According to our previous analysis of alkanes carried out with the use of the PALS technique, the best results of the spectrum decomposition are achieved assuming only one population of \emph{ortho}- and \emph{para}-Ps, whereas the ratio of the \emph{ortho} to \emph{para} intensity is fixed at 3/1.
\par
Tools of machine learning like genetic algorithms or artificial neural networks have been used to perform numerical calculations in a variety of aspects in positron science~\cite{Jegal22,Herraiz21,Jegal22,Wedrowski10,Whiteley20,Petschke19}. They have also been used for unfolding the lifetimes and intensities from PALS spectra~\cite{Lemes05,Pazsit99,An12,Viterbo01}. Possibly due to the low computing power of the hardware and the low time resolution of PALS spectrometers at the time when the neural network algorithms for decomposition of PALS spectra were proposed, most of the spectra used in these calculations are simulated by the software but not measured directly. For the same reason, the neural network architecture used there does not allow changing parameters as much as is allowed by algorithms developed today. Furthermore, no procedure allowing application for the same calculations of spectra registered for different time constants per channel has been presented since then. Thus, the preferred software used for spectrum decomposition is still based on non-linear fitting algorithms which do not include a possibility of establishing the result based on a multi-spectra set at the same time.
\par
\begin{figure}
\centering
\includegraphics[scale=0.2, angle=90]{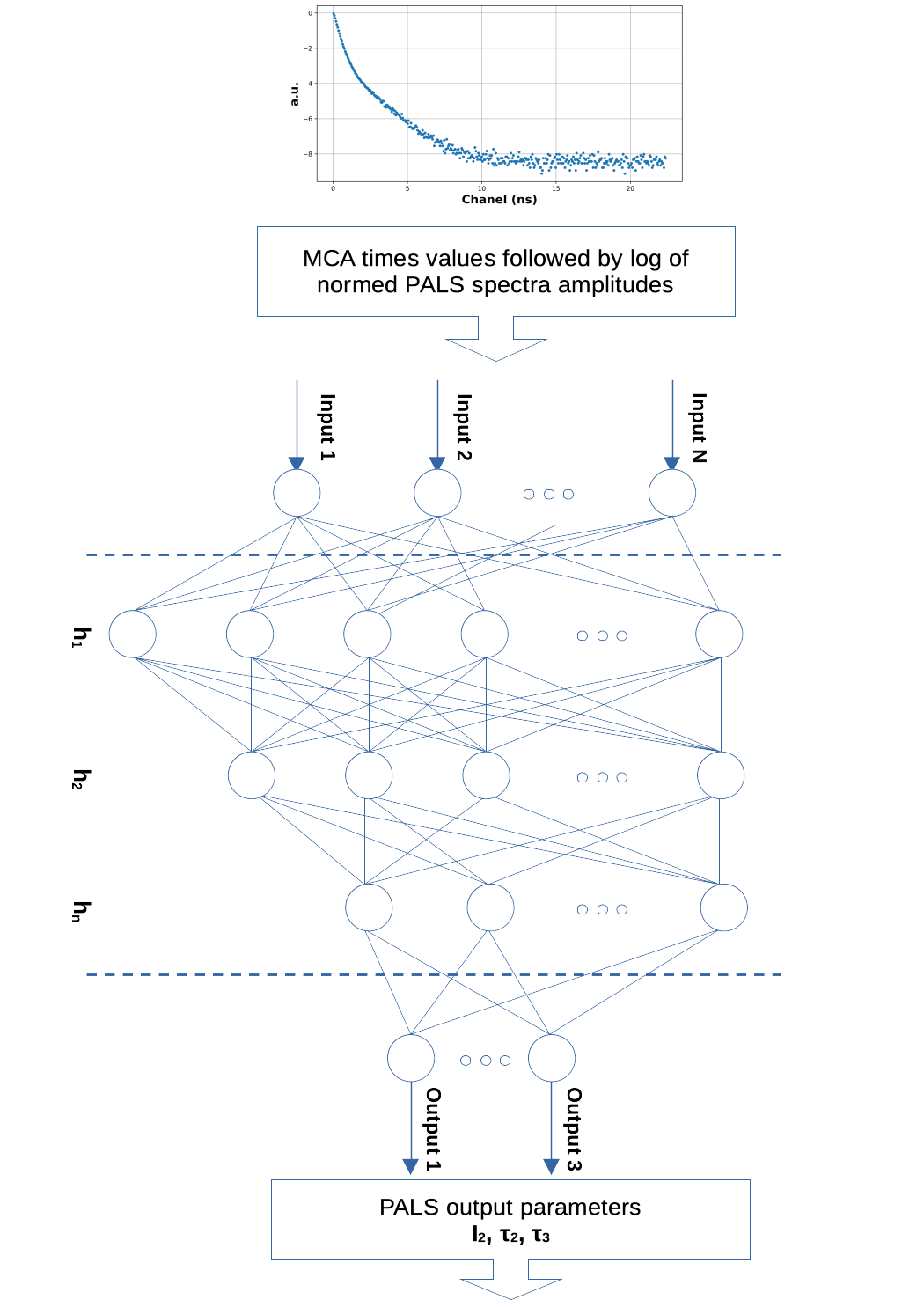}
\caption{Schematic view of the MLP applied. The PALS data from consecutive channels of the MCA are transferred as the amplitudes of the consecutive input neurons $\mathtt{In}_i$. $\mathtt{h}_i$ denote neurons in the $i$-th hidden layer whereas $\mathtt{Out}_i$ denote output neurons returning chosen PALS decomposition parameters.}
\label{fig:architecture}
\end{figure}

Here, we present an approach to analysis of PALS spectra based on the multi-layer perceptron (MLP) model, which is one of the tools of machine learning~\cite{Rebala19}. The model assumes a network of inter-connected neurons grouped in the input layer ($\mathtt{In}_i$), the hidden neurons layers ($\mathtt{h}_i^k$) and the output neurons layer ($\mathtt{Out}_i$), where $i$ goes over the neurons in a given layer and $k$ numbers the hidden layers. A graphical diagram of the network used is shown in fig.~\ref{fig:architecture}. The numbers of $\mathtt{In}$ and $\mathtt{Out}$ neurons are determined by the amount of the independent input data introduced to the network and the data defined to be the results of calculation in a given problem, respectively. The number of hidden layers and the number of neurons within these layers are set experimentally to optimise the network to give required results. To each layer (excluding the output layer), one bias neuron is attached for technical reasons~\cite{Heaton12}. The tool assumes the learning process first, where the $\mathtt{In}$ neurons are fed with the data for which the result of the $\mathtt{Out}$ neurons is known in advance. During this process, the weight coefficients for pairs of inter-connected neurons are adjusted by an algorithm, so that the output of the MLP can give results most similar to the expected ones. The MLP becomes to be trained after a number of iterations of training. Once the MLP results of learning are satisfied, the MLP can be used to calculate the output for the input data never used in the training process.
\par
The MLP type of network can be applied to solve the problem of both classification and regression. For the first group of problems, it is required from the MLP to ascribe the values of the output parameters in the form of well separated categories. These so-called \emph{labels} can always be parametrised by a discrete set of numbers. The problem described in this paper is classified as rather a regression problem (MLPR) where the values of the output at each $\mathtt{Out}$ neuron are characterised by a continuous set of values. Consequently, the output may contain values approaching these appearing during the learning process but may not necessarily be exactly of the same value. The internal algorithms of the MLPR allows regarding the learning process as a way of finding the quasi-continuous output function of input parameters. In our case, based on the data from the PALS spectra applied as the input values of the perceptron, the MLPR is used for solving the regression problem of finding the values of key PALS parameters on the output.
\section{Method}
The \emph{scikit-learn} library was used to estimate the PALS parameters for alkanes \cite{Abraham11}. In our case, the MLP regressor class (called \emph{MLPRegressor}), which belongs to one of the supervised neural network models, was implemented. In this class, the output is a set of continuous values. It uses the square error as the loss function. This model optimises the squared error using the Broyden–Fletcher–Goldfarb–Shanno algorithm (LBFGS)~\cite{LBFGS}, which belongs to quasi-Newton methods. Some MLPRegressor parameters playing a key role are mentioned below. Their values require to be tuned, especially the $\mathtt{alpha}$ hyper-parameter, which helps in avoiding over-fitting by penalising weights with large magnitudes. A full list of parameters of MLPRegressor is defined in~\cite{scikit}.
\par
In the learning process here, we used spectra collected for years from an analog spectrometer for several alkanes (in the range of C$_6$ -- C$_{40}$) measured at several temperatures (-142$^{\circ}$C -- 100$^{\circ}$C). Irrespective of both the goal of the particular experiment and the length of the alkane chain used as a sample, the initial assumptions made for starting the analysis of the spectra made by the LT programme~\cite{Kansy96} were the same. Each measurement resulting in the spectra used was performed with a sample prepared in a similar way, i.e. the sample was degassed, and the rate of cooling or heating was the same. In each case, the measurement at constant temperature took place for at least one hour which gave some hundreds of thousands of annihilations (the strength of the radioactive source was similar in each case). During some experiments the temperature was changed stepswise but each spectrum was collected at constant temperature. The most important issue here is that the post-experimental analysis of the spectra was conducted under the same general assumptions every time. Especially, for the decomposition of these spectra, we used LT supposing that the time resolution curve can be approximated by one-gaussian curve. Every time it was assumed that the annihilation process in the Kapton envelope accounted for 10\% (so-called \emph{source correction}). Additionally, only one component was always assumed for \emph{para}- and \emph{ortho}-Ps, whereas their intensity ratio was fixed at the value 3/1 (see~\cite{Goworek09} for details of the experimental procedure).
\par
Taking into account these assumptions, each spectrum was decomposed into three exponential curves for which the intensities ($I$) and lifetimes ($\tau$) were calculated for the following sub-populations of positrons: the free positron annihilation ($I_2$, $\tau_2$), \emph{para} ($I_1$, $\tau_1$), and \emph{ortho}-Ps ($I_3$, $\tau_3$)\footnote{Numbering of the indices is related to the length of $\tau$. The increasing values of the indices correspond to the rising length of lifetime.}. The database collected in this way contained 7973 PALS spectra, wherein about 75\% were used in the neural network training process and the rest were used as a testing set for checking the accuracy of the results given by the learned network.
\par
The number of input neurons is determined by the number of channels of the Multi-channel Analyser (MCA) module of the PALS spectrometer recording PALS spectra. Furthermore, the number of the output neurons is related in this model to the number of PALS parameters, which are supposed to be predicted for further studies of physical processes in the sample. The decomposition of the PALS spectrum made by commonly used programs, like LT, allows determining ($I$,$\tau$) pairs for all assumed components of a given spectrum. However, often, not all these parameters are needed for further analysis. Furthermore, some of these parameters are inter-dependent. For example, in the case of PALS spectra for the alkanes discussed here, one assumes that the spectrum is built up by events from the three populations of positrons mentioned above ($\tau_1$ -- $\tau_3$, $I_1$ -- $I_3$ parameters). However, from the practical view point, only $\tau_2$, $I_2$, $\tau_3$, and $I_3$ are then used for studying physical processes and the structure of the sample. Furthermore, in this case, $I_i$ are inter-dependent and fulfil the following relations $I_1$+$I_2$+$I_3$=100\%\footnote{Annihilation in Kapton was subtracted in advance.} and $I_3/I_1$=3. Thus, effectively, the parameters considered as the $\mathtt{Out}$ parameters of MLPR are only $I_2$, $\tau_2$, and $\tau_3$. According to this, we declared in our modelling only three output neurons for receiving values for these three parameters.
\section{Preparation of input and output data}
\label{sec:Preparation}
\begin{figure}
\centering
\includegraphics[scale=0.5]{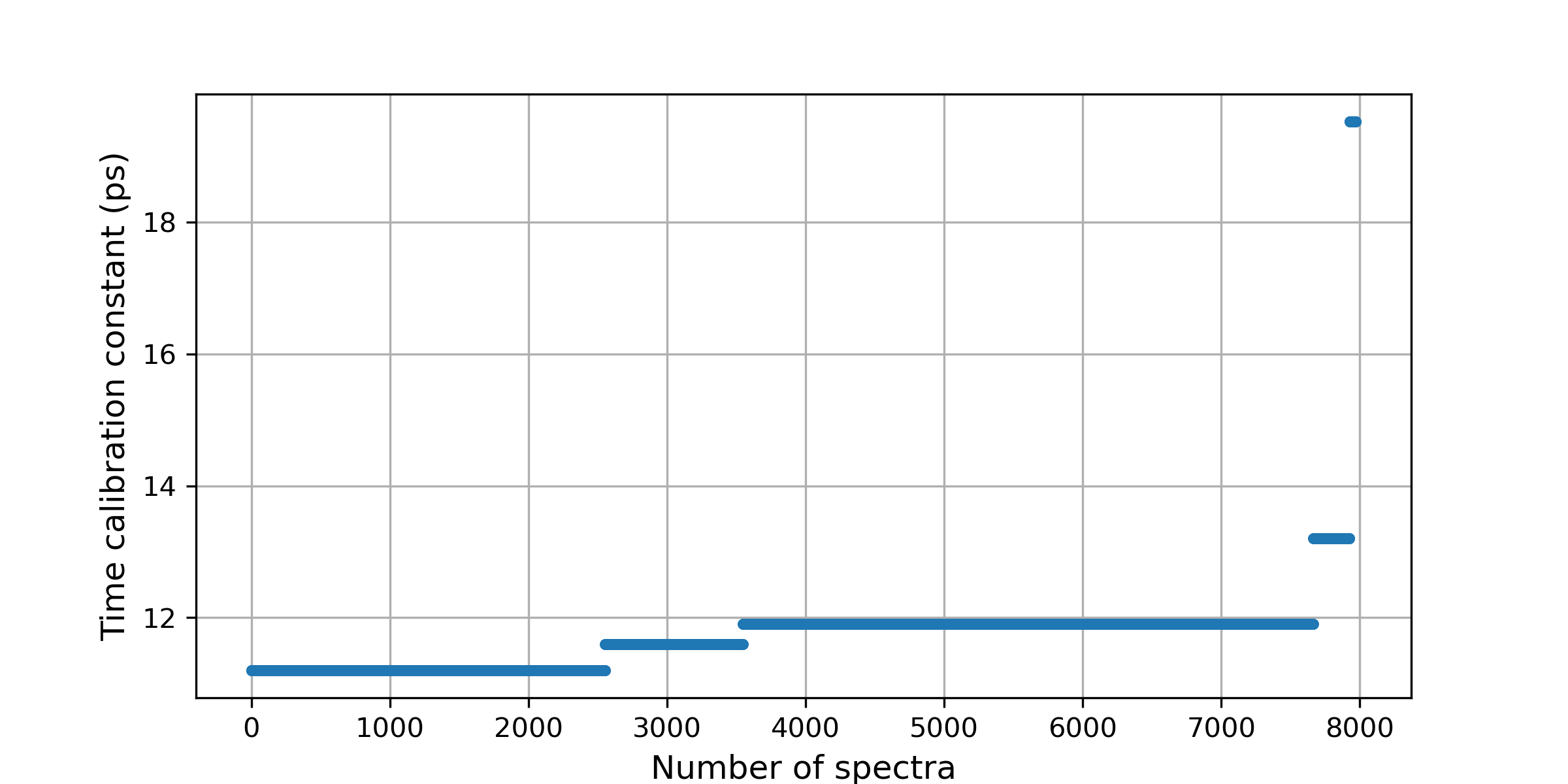}
\caption{Number of spectra (horizontal axis) with a given value of the time constant per channel $\Delta$ (vertical axis) used as a data set in the presented calculations.}
\label{fig:TimeConstantsSet}
\end{figure}

During the PALS measurements, the time constant per channel ($\Delta$) varied, depending on the internal properties and settings of the spectrometer. Most of the data used here were collected with $\Delta$=11.9~ps; however, some spectra were measured with $\Delta$=11.2~ps, 13.2~ps, 11.6~ps, and 19.5~ps (fig.~\ref{fig:TimeConstantsSet}).
Therefore, it is important for the $\mathtt{In}$ neurons to code the PALS amplitude samples not in the relation to the channel numbers but in the scale of time. Hence, in addition to the spectrum amplitudes, the regressor has to learn the times associated with these amplitudes. Thus, one half of the $\mathtt{In}$ neurons is fed with time values for consecutive channels of a spectrum, whereas the second half is fed with the values of their amplitude. The advantage of the regression approach applied here is the ability to test spectra measured even for a time sequence that has never appeared in an extreme case in the training process.
\par
\begin{figure}
\centering
\includegraphics[scale=0.5]{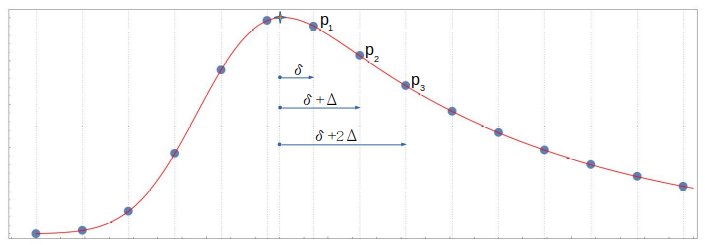}
\caption{Schematic view of a peak region of the PALS spectrum. The bullets indicate $(t,log(A))$ pairs saved in the MCA channels, whereas the star indicates a position of a peak calculated assuming a gaussian shape of an apparatus distribution function. Only the points to the right of the star ($p_1$, $p_2$, ...) are taken as data introduced to MLPR. The $\delta$ parameter denotes a time distance between the calculated peak and the first point, whereas $\Delta$ is the time distance between two points.}
\label{fig:normalization}
\end{figure}

This method requires setting correctly a common zero-time for each spectrum. To achieve this, the original data from the left slope of the spectrum peak (and only a few points to its right) were used to interpolate the resolution curve, which is assumed to be in the gaussian form. The position of this peak defines a zero-time for a spectrum. One-gaussian interpolation is compatible with previous LT analysis assumptions. Based on the common starting position for all spectra established in this way, the values of time for each channel on the right to the peak were re-calibrated for each spectrum depending on $\Delta$ for which the spectrum was measured. Finally, for further analysis, we took the same $N$ number of consecutive channels for each spectrum on the right to its peak (points $p_i$ in fig.~\ref{fig:normalization}). The $\delta$ parameter shown in fig.~\ref{fig:normalization} denotes the distance (in time units) between the first point on the right to the peak and the calculated time position of the peak. The number $N$ taken for further analysis was established experimentally. Finally, the spectrum data for the MLPR input are the $N$ points $p_i$ with their two values: the re-calibrated number of counts in a given channel (see below) and their re-calibrated times of annihilation.
\par
\begin{figure}
\centering
\includegraphics[scale=0.5]{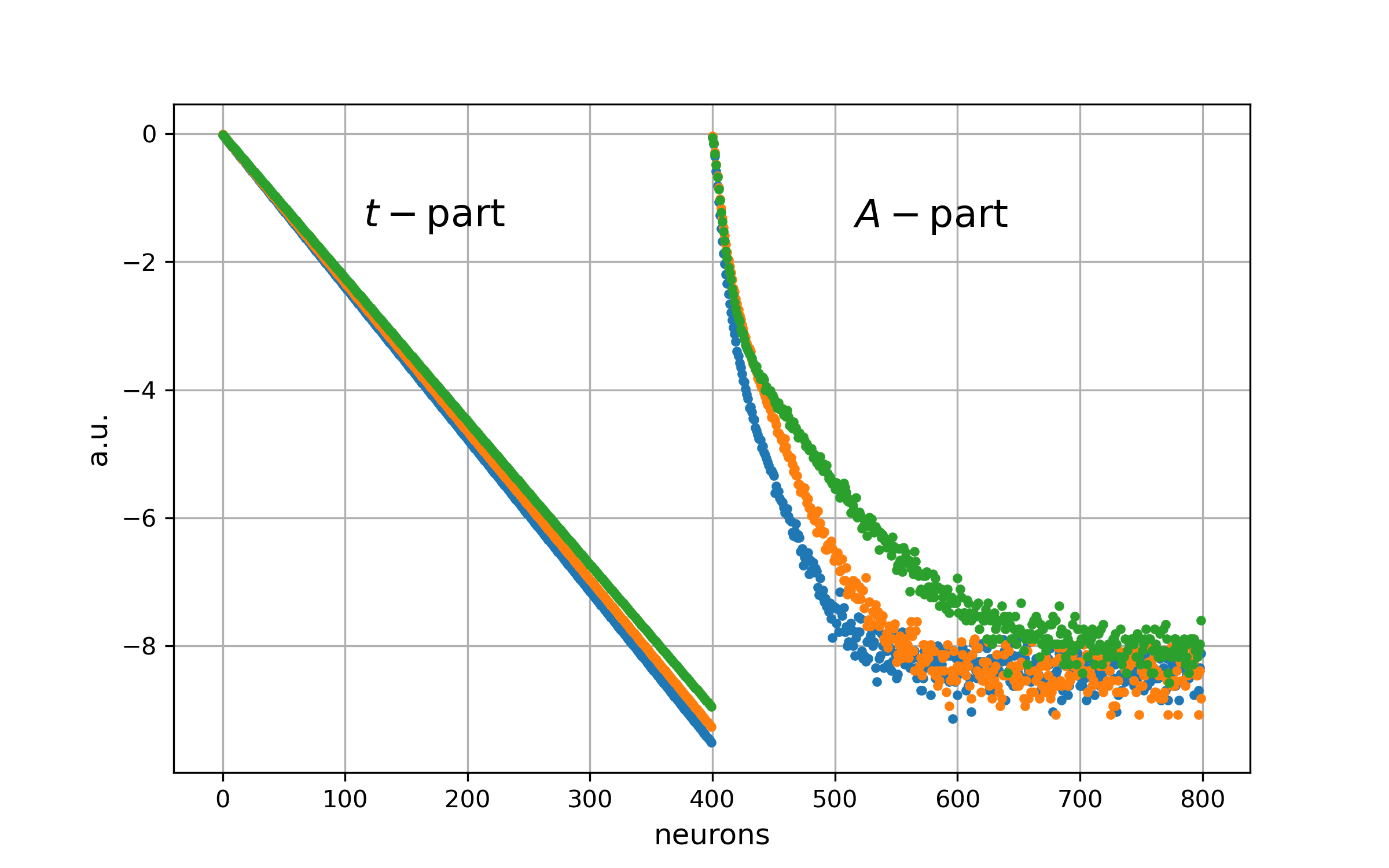}
\caption{Input data directed to $\mathtt{In}$ neurons can be divided into two sub-sets: $t$-part which is a set of time values for points $p_1$, $p_2$, ... (see fig.~\ref{fig:normalization}) and $A$-part coding the log function of their normalised amplitudes. In special cases, these data are smoothed or compressed before use in MLPR.}
\label{fig:InputSpectra}
\end{figure}

Then, to minimise errors, the original input data were transformed before application. Each original spectrum was stored in 8192 channels of MCA. Firstly, starting from the first channel on the right to the spectrum maximum ($p_1$ in fig.~\ref{fig:normalization}), 2k channels were taken from the original spectrum. This means that the spectra were truncated at about 25~ns of the registration time (varying to some extent, depending on the $\Delta$ for a given spectrum). Secondly, to smooth random fluctuations, the data were smoothed in most cases. One of the examples of smoothing is averaging over five consecutive channels. In this case, the number of samples in each spectrum shrank from the original 2k channels to the amount of 400. Since the $\mathtt{In}$ neurons transfer information about the pair of values -- times ($t$-part) and amplitudes ($A$-part), 800 input neurons that fed the MLPR with the data in this case were declared. Thirdly, to standardise the range of the input data values, the set of the PALS amplitudes was normalised to the maximum value of the amplitude and then logarithmised. According to these transformations, the $A$-part data covered the numerical range [-9,0] -- fig.~\ref{fig:InputSpectra}. Furthermore, to adjust the range of the values in the $t$-sector, the values of time were divided by -2.5. As a result, all data transferred to the $\mathtt{In}$ neurons were in the range of [-10,0].
\par
Additionally, we applied some transformation of the original values for the $\mathtt{Out}$ neurons in order to have their values at each neuron scaled to the same range. Initially, the first output neuron is related to $I_2$, whereas its original value range is typically tenths (in \% units). The second neuron transfers the information related to $\tau_2$ whose original values are of the order of 0.1 (of ns), whereas the order of $\tau_3$ related to the third neuron is originally 1 (of ns). In order to have the uniform order of numerical values on all $\mathtt{Out}$ neurons, the data that finally feed with them are [$I_2$/10, $\tau_2 \cdot$10, $\tau_3$].
\par
The criterion of acceptance of training the network was the best value of the score validation function defined for this regressor as
\begin{equation}
\mathcal{S}=1-\frac{\sum_{N}(\mathcal{O}_{\text{true}}-\mathcal{O}_{\text{pred}})^2}{\sum_{N}\mathcal{O}_{\text{true}}^2},
\end{equation}
where $\mathcal{O}_{\text{true}}$, $\mathcal{O}_{\text{pred}}$ -- expected (known) and calculated (predicted) values of the result, respectively~\cite{Abraham11,scikit}. $\mathcal{S}$ is calculated for both the learning and testing sets separately. $N$ here denotes the number of spectra in the trained or tested set. The optimum value of $\mathcal{S}$ is~$\approx$1.
\section{Results}
The MLPRegressor used in these calculations requires establishing some key parameters~\cite{scikit} influencing the ability to learn and a speed of the learning process. We performed some tests trying to optimise these parameters. The best results we obtained by the settings shown in tab.~\ref{tab:params}.
\begin{table}[htbp]
\centering
\caption{Values of the MLPRegressor parameters applied for producing the final MPLR results.}
\begin{tabular}{|r|c|}
\hline
Parameter & value\\ \hline\hline
$\mathtt{hidden\_layer\_sizes}$ = & 7$\times$150\\ 
$\mathtt{activation}$ = & \emph{relu}\\
$\mathtt{solver}$ = & \emph{lbfgs} \\
$\mathtt{alpha}$ = & 0.01 \\
$\mathtt{learning\_rate}$ = & \emph{invscaling}\\
$\mathtt{power\_t}$ = & 0.5 \\ 
$\mathtt{max\_iter}$ = & 5e+9 \\ 
$\mathtt{random\_state}$ = & None \\ 
$\mathtt{tol}$ = & 0.0001 \\
$\mathtt{warm\_start}$ = & True \\ 
$\mathtt{max\_fun}$ = & 15000 \\
\hline
\end{tabular}
\label{tab:params}
\end{table}
Both the names and the meaning of the technical parameters shown in the table are identical to these defined in the routine description~\cite{scikit}. Once the key parameters of the MLPR were established (especially the $\mathtt{solver}$), we performed tests of credibility of the network changing the number of hidden layers, the number of neurons within ($\mathtt{hidden\_layer\_sizes}$ parameter), and the $\mathtt{alpha}$ parameter. The results in tab.~\ref{tab:DifferentNetworks} show examples of the results. For these networks, we specified the mean validation score parameter for both the training $\langle\mathcal{S}_{tr}\rangle$ and testing $\langle\mathcal{S}_{te}\rangle$ sets separately with their variation $\delta\mathcal{S}$. Averaging was made over the results of ten runs of the training process for identical networks differing by initially random weights. We did not notice any rule giving a ratio of the numbers of neurons that should be declared in the consecutive hidden layers (especially as the number of neurons should decrease proportionally in the consecutive layers). A few initial examples shown here suggest that the accuracy of results increases when both the number of hidden layers and the number of neurons inside increase. However, the last two rows of the table show that a further increase in these parameters does not give better results. Finally, the network that gave a nearly best result was chosen (marked in bold $\langle\mathcal{S}_{tr}\rangle$ in the table). It was checked for this network that an increase in the iterations of training ($\mathtt{max\_iter}$ parameter) beyond about 5$\cdot$10$^{9}$ did not improve $\langle\mathcal{S}\rangle$.


\begin{table}[htbp]
\centering
\caption{Valuation score for chosen values of some MLPR parameters. $\mathcal{S}$ values are averages over 10 runs with random initial neuron weights. A nearly optimum case of parameters is placed in a row with $\mathcal{S}$ marked in bold.}
\begin{tabular}{|c|c|c|c|c||c|c|c|c|}
\hline
$\mathtt{hidden\_layer\_sizes}$ & $\mathtt{max\_iter}$ & $\mathtt{alpha}$ &  $\langle\mathcal{S}_{tr}\rangle$ & $\delta\mathcal{S}_{tr}$ & $\langle\mathcal{S}_{te}\rangle$ & $\delta\mathcal{S}_{te}$\\
\hline\hline
$30 \times 25 \times 15$ & 10$^6$ & 0.7 & 0.950 & 0.003 & 0.942 & 0.004\\
\hline
$3 \times 100$ & 10$^8$ & 0.7  & 0.969 & 0.005 & 0.965 & 0.008\\
\hline
$3 \times 100$ & 10$^8$ & 0.1  & 0.974 & 0.005 & 0.968 & 0.008\\
\hline
$3 \times 100$ & 5$\cdot$ 10$^8$ & 0.1  & 0.975 & 0.003 & 0.976 & 0.003\\
\hline
$4 \times 100$ & 10$^8$ & 0.1  & 0.978 & 0.004 & 0.975 & 0.006\\
\hline
$500 \times 400 \times 300 \times 200$ & 5$\cdot$10$^9$ & 0.01  & 0.977 & 0.003 & 0.974 & 0.007\\
\hline
$7 \times 150$ & 5$\cdot$10$^8$ & 0.01  & \textbf{0.985} & 0.002 & 0.975 & 0.013\\
\hline
$\begin{array} {rl} & 500 \times 500 \times 400 \times 400 \times \\ & \times 300 \times 300 \times 200 \times 200 \end{array}$ & 5$\cdot$10$^9$ & 0.01  & 0.978 & 0.005 & 0.977 & 0.008\\
\hline
$8 \times 500$ & 5$\cdot$10$^9$ & 0.01  & 0.982 & 0.004 & 0.980 & 0.005\\
\hline
\end{tabular}
\label{tab:DifferentNetworks}
\end{table}
\par
\begin{figure}[htbp]
\centering
\includegraphics[scale=0.3]{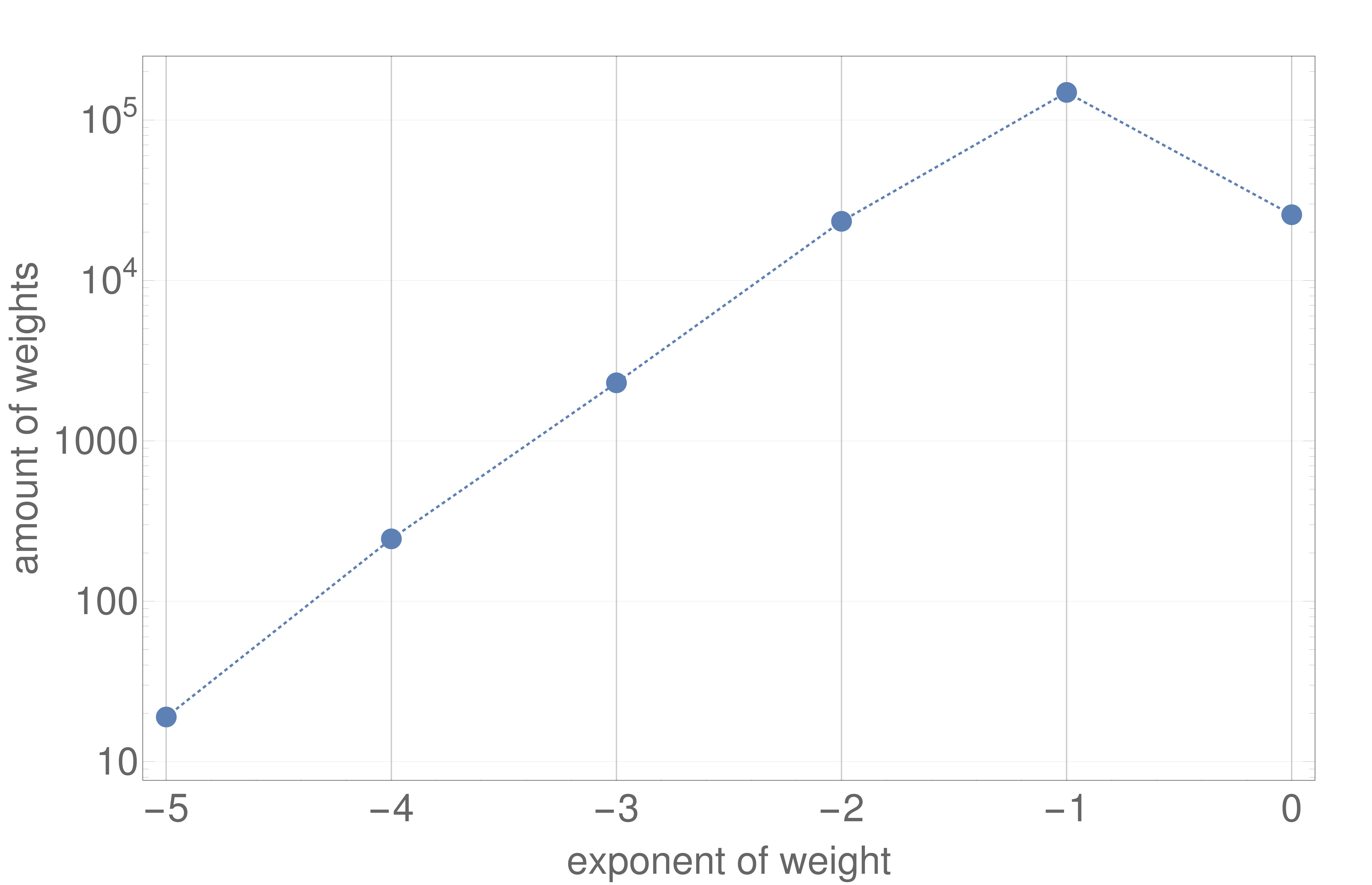}
\caption{Number of cases (log scale) of exponents of weights for a network with 7$\times$100 hidden layers. For this network, the key parameters are: $\mathtt{solver}$=\emph{lbfgs}, $\mathtt{max\_iter}$=5$\cdot$10$^{11}$, $\mathtt{alpha}$=0.005, $\mathtt{learning\_rate}$=\emph{invscaling}, and $\mathtt{activation}$=\emph{relu}.}
\label{fig:weights}
\end{figure}

For several finally tested networks, the spectrum of the magnitude of inter-neurons weights was checked. It is expected that weights that differ significantly from the average range of values may affect the stability of the results. In this case, the range of weight values seems to be quite narrow. As shown in fig.~\ref{fig:weights}, the weight magnitude order (exponent of weights) for the chosen network ranges from 10$^{-5}$ to 10$^{0}$, while the relative number of cases in these subsets changes exponentially. The lack of values outside the narrow set of values suggests that self-cleaning of the resultant weights is performed by the MLPRegressor algorithm itself.

%
\par
The number of all PALS spectra used as a database for the network was 7973, and 6500 were used to learn the output values (training set) by the network, while the rest were used for checking the results of learning (testing set). Tab.~\ref{tab:exampleA} shows a few examples of randomly taken results given by one of the networks finally used.
\begin{table}[htbp]
\centering
\caption{Examples of a few randomly taken results of calculations (prediction) of the $I_2$, $\tau_2$, and $\tau_3$ parameters compared to the expected values calculated by LT. Here, $\mathtt{hidden\_layer\_size}$=7$\times$150, $\mathcal{S}$=0.985 for both training and testing sets.}
  \begin{tabular}{|c|c|c|c|c|c|c|}
\hline
    \multirow{2}{*}{Example} &
      \multicolumn{2}{c|}{$I_2$ [\%]} &
      \multicolumn{2}{c|}{$\tau_2$ [ns]} &
      \multicolumn{2}{c|}{$\tau_3$ [ns]} \\
    & \scriptsize{expected} & \scriptsize{predicted} & \scriptsize{expected} & \scriptsize{predicted} & \scriptsize{expected} & \scriptsize{predicted} \\
\hline
    \scriptsize{1} & 68.0 & 68.8 & 0.27 & 0.28 & 1.21 & 1.20 \\
    \hline
    \scriptsize{2} & 47.2 & 47.6 & 0.38 & 0.39 & 3.23 & 3.18 \\
    \hline
    \scriptsize{3} & 65.4 & 65.5 & 0.30 & 0.29 & 1.35 & 1.38 \\
	\hline
    \scriptsize{4} & 78.3 & 78.3 & 0.23 & 0.24 & 1.11 & 1.12 \\
	\hline
    \scriptsize{5} & 52.4 & 52.4 & 0.35 & 0.34 & 2.93 & 2.93 \\
	\hline
    \scriptsize{6} & 60.5 & 60.4 & 0.31 & 0.30 & 1.25 & 1.20 \\
	\hline
    \scriptsize{7} & 59.3 & 58.8 & 0.29 & 0.29 & 1.19 & 1.22 \\
	\hline
    \scriptsize{8} & 61.0 & 62.3 & 0.30 & 0.31 & 1.91 & 1.91 \\
	\hline
    \scriptsize{9} & 70.2 & 69.5 & 0.21 & 0.21 & 1.06 & 1.10 \\
	\hline
    \scriptsize{10} & 39.9 & 38.8 & 0.23 & 0.22 & 1.15 & 1.13 \\
\hline
  \end{tabular}
\label{tab:exampleA}
\end{table}
The results given by the trained network were compared to the expected values known from the LT analysis. Although $\mathcal{S}$ for both the trained and tested sets in this case is not the highest one obtained in our tests, the result of the use of this network is satisfactory in a practical sense because the deviation of the predicted and expected result is in the range of deviation given by LT itself.
\par
The problem of pre-preparation of spectra for calculations by MLPR is worth mentioning. The main problems are where the spectrum should be cut and to what extent it is acceptable to smooth the spectra by averaging their consecutive values. As for the first problem, it was determined by series of runs for which the spectra were cut at other that mentioned limit of 2k channels that this number of channels was almost the best choice. $\langle \mathcal{S}\rangle$ was found to worsen in the case of a shorter cut (say, 1.5k channels), and did not improve significantly in the case of the longer ones (e.g. 3k channels) (but it took longer to compute the result because of an increase in the number of $\mathtt{In}$ neurons).
\par
The accuracy of the prediction increases when the learning and testing processes are limited to one only $\Delta$ with all parameters of the network kept constant. In this case, the set of values in the $t$-part for every spectrum varies in a much narrower range (only $\delta$ changes). In this case, the training process is more effective even if the size of the training set is reduced. To show this, we separated the set of spectra measured for only one $\Delta$=11.9~ps. Consequently, the whole set of samples under consideration shrank to 4116 members, and 3000 of them were used for training after the transformations described above. The score $\mathcal{S}$ obtained in this case was much greater than $\mathcal{S}$ for an identical network applied to spectra with all possible $\Delta$. The comparison of these two cases is shown in tab.~\ref{tab:DeltaDiff} (the last row of the table). Here, to have the result reliable for networks with different (random) initial weights, the score was averaged over 30 runs.
\begin{table}[htbp]
\caption{Comparison of MLPR validation score $\mathcal{S}$ for different formats of the input data. Comparison of the results for 'raw' data (log of normalised and adjusted data according to the procedure described in section~\ref{sec:Preparation}) and data on which the moving average and compressing average (by each 3 and 5 separate spectrum points) are applied. The result for the network fed with the data collected for one chosen $\Delta$ is added in the last row of the table.}
\centering
\begin{tabular}{|llc|c|c|}
\hline
\multicolumn{3}{|c|}{MLPR and spectra parameters}                                                                                                                                                                                                                                                                                                                                                                                                         & $\langle\mathcal{S}_{tr}\rangle$                 & $\langle\mathcal{S}_{te}\rangle$    \\ \hline
\multicolumn{1}{|l|}{\multirow{7}{*}{\begin{tabular}[c]{@{}l@{}}$\mathtt{solver}$=\emph{lbfgs}\\ $\mathtt{activation}$=\emph{relu}\\ $\mathtt{learning\_rate}$=\emph{invscaling}\\ $\mathtt{alpha}$=0.01, $\mathtt{In}$=800\\ $\mathtt{hidden\_layer\_sizes}$=7$\times$150\\$\mathtt{max\_iter}$=5$\times$10$^9$\\ averaged over 30 trials\end{tabular}}} & \multicolumn{1}{l|}{\multirow{4}{*}{\begin{tabular}[c]{@{}l@{}}Several $\Delta$s \\ $N_{tr}$=6500 ($\sim$ 80\%)\\ $N_{te}$=1473\end{tabular}}}       & unsmoothed data             &     0.984$\pm$0.005                  &    0.963$\pm$0.006                   \\ \cline{3-5} 
\multicolumn{1}{|l|}{}                                                                                                                                                                     & \multicolumn{1}{l|}{}                                                                                    & moving average             & 0.984$\pm$0.005            & 0.977$\pm$0.007             \\ \cline{3-5} 
\multicolumn{1}{|l|}{}                                                                                                                                                                    & \multicolumn{1}{l|}{}                                                                                                     & k=3                  & 0.981$\pm$0.005    & 0.976$\pm$0.007    \\ \cline{3-5} 
\multicolumn{1}{|l|}{}                                                                                                                                                                                                                            & \multicolumn{1}{l|}{}                                                                                & k=5                  & 0.981$\pm$0.006        & 0.974$\pm$0.010  \\ \cline{2-5} 
\multicolumn{1}{|l|}{}                                                                                                                                                                                                                            & \multicolumn{1}{l|}{\multirow{3}{*}{\begin{tabular}[c]{@{}l@{}}Fixed $\Delta$=11.9 ps\\ $N_{tr}$=3000 ($\sim$73\%)\\ $N_{te}$=1116\end{tabular}}} & \multirow{3}{*}{k=5} & \multirow{3}{*}{0.993$\pm$0.002} & \multirow{3}{*}{0.989$\pm$0.003} \\
\multicolumn{1}{|l|}{}                                                                                                                                                                                                                            & \multicolumn{1}{l|}{}                                                                                                                                                         &                      &                    &                     \\
\multicolumn{1}{|l|}{}                                                                                                                                                                                                                            & \multicolumn{1}{l|}{}                                                                                                                                                         &                      &                    &                     \\ \hline
\end{tabular}
\label{tab:DeltaDiff}
\end{table}
\par
The validation score $\mathcal{S}$ is sensitive to smoothing the spectrum which reduces to some extent the information given by the PALS spectrum. In tab.~\ref{tab:DeltaDiff}, two cases are compared where each 3- and 5-tuples of points of the spectrum (forming non-overlapping windows) were taken to calculate their average amplitude. For example, $N$=3500 points of the initial spectrum are reduced to 700 points when averaging over $k$=5 points; when the remainder of the division of $N$ by $k$ is not zero, an integer quotient is taken. $\langle \mathcal{S}\rangle$ calculated for these two cases shows that both of them give the same results statistically. However, further shrinking the spectrum by setting $k$=6 or more produces worse $\mathcal{S}$.
\par
Tab.~\ref{tab:DeltaDiff} also shows the $\mathcal{S}$ parameter when the moving average is applied during preparation of spectra. The sampling window applied here is 10. The comparison of this result to the result of calculation with unsmoothed data shows that the application of the moving average does improve predictions for the testing data set.
\section{Conclusions}
We have shown in this paper that the easy-to-reach machine learning MLPRegressor tool enhanced with some programming in \emph{Python} making some preparation of data, can be used as an alternative method of solving the problem of inversion of PALS spectra. The main disadvantage of the presented method is the need of decomposition of training spectra by other software to have $\mathtt{Out}$ values for training. Once the training set is collected and the network is trained, the algorithm works very quickly, giving the result for the tested spectrum. The training process used here is based on results given by LT, i.e. a method producing results with some uncertainty itself. The uncertainty produced by the LT is caused by the use of numerical methods to compute the fit in particular cases. On the other hand, since the MLPR prediction bases on information from a large set of spectra, this approach seems to be less sensitive to the specific shape of a given spectrum and may be more accurate in predicting parameters. Furthermore, the presented method seems to be faster than the referenced ones, since calculations made by a trained network are reduced to simple transformations of matrices and vectors, which is not demanding computationally and less sensitive to numerical problems.
\par
Although the model presented here is similar to that described in~\cite{Pazsit99} (and repeated in~\cite{An12}), there are significant differences indicated in tab.~\ref{tab:PazsitVsOur}.
\begin{table}
\centering
\caption{Comparison of key parameters and results of the MLPR modelling applied in this study (\emph{skLearn}) and a three-component spectrum analysis published previously (presented in~\cite{Pazsit99} and~\cite{An12}).}
\begin{tabularx}{\textwidth}{|X|c|c|c|}\hline
 & \textbf{P\'{a}zsit}~\cite{Pazsit99} & \textbf{An}~\cite{An12} & \textbf{skLearn}\\
\hline
Type of training spectra                                                              & simulated & simulated & real (alkanes)\\ 
No. of training spectra                                                               & 575 & 920 & 7973\\ 
Type of spectra tested                                                                & simulated & simulated, silicon & alkanes \\ 
No. of test spectra                                                                   & 50 & 100 (30) & 1473\\ 
Type of network                                                                       & one-layer perc. & one-layer perc. & multi-layer perc.\\ 
Channel width [ps]                                                                    & 23.2 & 24.5 & \textbf{some} (11.2-19.5)\\ 
No. of MCA/taken channels                                                             & 1500/1500 & 1024/1024 & 8192/3500\\ 
Approx. no. of counts in spec.                                                        & 10M & 10M & $\sim$400k\\ 
Solver                                                                                & backward error prop. & backward error prop. & \textbf{some}\\ 
No. of hidden layers                                                                  & 1 & 1 & some\\ \hline
$I_2$, $\tau_3$ average error [\%] \newline on tested \textbf{simulated} spectra                & 7.3, 1.0 & 1.07-3.52, 0.55-1.21 & -, -\\ 
$I_2$, $\tau_3$ average error [\%] \newline on tested \textbf{real} spectra                     & -, - & -, - & 1.03, 1.70 \\ \hline
\end{tabularx}
\label{tab:PazsitVsOur}
\end{table}
Our experimental data are collected by spectrometers differing in functional properties, especially differing in time resolution. Even for one spectrometer, this parameter should be re-calibrated periodically due to changes in experimental conditions, especially temperature. In the algorithm presented in~\cite{Pazsit99}, the same resolution curve for all spectra is assumed. In our data preparation procedure, the parameters of the resolution curve are interpolated for each case. Based on this, the $\delta$ parameter is calculated and the value of the shift in time is established for consecutive channels. Although one-gaussian resolution curve was assumed here, it is possible to extend this algorithm for much more complicated cases where the distribution curve consisted in a sum of gaussians, for example. As already mentioned in~\cite{Pazsit99}, in that case, a possibility of recognising a distribution function would give compatibility to MELT~\cite{Shukla93}. Such an extension requires extending the calculations by applying another neural network, working in advance, which returns the parameters of the resolution curve in a given case. This problem has been solved by application of a Hopfield neural network~\cite{Viterbo01}. Taking into account our collection of spectra, it was checked with the use of LT and (occasionally) with MELT that the apparatus resolution curve is one-gaussian for our spectra. Hence, they do not allow testing such an extended model.
\par
Furthermore, the MCA module of spectrometers may differ in the time constant per channel $\Delta$. Thus, spectra used as a training data set may be collected for different channel widths. Taking into account the method presented in~\cite{Pazsit99} for fixed $\Delta$, the training result is of little use for spectra collected with another $\Delta$. Oppositely, we have shown the possibility of application of an improved algorithm to data collected for different $\Delta$s. The data collected from many spectrometers may contribute to a large training data set, which allows solving the inversion problem for any PALS spectrum and, thus, may be a universal tool that can be used in different laboratories. Although the set of $\Delta$ used here is small, the accuracy of the results is quite good. To use this tool to determine the real-world spectrum parameters, the training process should be extended by adding the spectra measured for a wider range of $\Delta$.
\par
For greater generalisation, it is possible in principle to attach spectra collected for other compounds to the training data set. For consistency, it suffices for a training database to keep the same number of components (three here) in spectrum decomposition. However, in practice, some incompatibilities of the spectra for different compounds may arise because decomposition into a few exponential processes is probably always a simplification of a real case where some distribution of the size and shape of free volumes should be taken into account as well as other Ps formation details.
\par
Although the approach presented here is reduced to the analysis of alkanes solely, the algorithm can be applied in calculation of PALS parameters of other types of samples as well.
\bibliographystyle{unsrt}
\bibliography{bib}
\end{document}